\documentclass[11pt,a4paper]{article}
\pdfoutput=1
\usepackage{jcappub}

\usepackage{xcolor}

\usepackage{mathrsfs}
\usepackage{bm}
\usepackage{subfigure}
\usepackage{feynmp}
\usepackage{extarrows}
\usepackage{slashed}
\usepackage{dcolumn}
\usepackage{verbatim}
\usepackage{here}
\usepackage{multirow} 
\allowdisplaybreaks[4]

\numberwithin{equation}{section}

\newcommand{\fr}[2]{\mbox{$\frac{\,{#1}\,}{#2}$}}

\newcommand{\nn}{\nonumber}
\def\bge{\begin{equation}}
\def\ede{\end{equation}}
\def\bga{\begin{aligned}}
\def\eda{\end{aligned}}
\def\bgp{\begin{pmatrix}}
\def\edp{\end{pmatrix}}
\def\bgs{\begin{subequations}}
\def\eds{\end{subequations}}
\newcommand{\beq}{\begin{equation}}
\newcommand{\eeq}{\end{equation}}
\newcommand{\bq}{\begin{equation}}
\newcommand{\eq}{\end{equation}}
\newcommand{\ba}{\begin{array}}
\newcommand{\ea}{\end{array}}
\newcommand{\beqa}{\begin{eqnarray}}
\newcommand{\eeqa}{\end{eqnarray}}
\newcommand{\beqs}{\begin{subequations}}
\newcommand{\eeqs}{\end{subequations}}

\def\[{\left[}
\def\]{\right]}
\def\({\left(}
\def\){\right)}

\def\dis{\displaystyle}

\def\leqq{\leqslant}
\def\geqq{\geqslant}

\def\leqq{\leqslant}
\def\geqq{\geqslant}

\def\Phit{\Phi_{\text{tot}}^{}}

\def\to{\rightarrow}

\def\End{\end{document}}

\title{\huge Constraining Astrophysical Neutrino \\ Flavor Composition from Leptonic Unitarity}

\author[a]{\large Xun-Jie Xu,}
\author[a]{\large~ Hong-Jian He,}
\author[b]{\large~ Werner Rodejohann\,}

\affiliation[a]{Institute of Modern Physics and Center for High Energy Physics,\\
                Tsinghua University, Beijing 100084, China.\\[1mm]
                Center for High Energy Physics, Peking University, Beijing 100871, China.
                \\[-4mm]
                }
\affiliation[b]{Max-Planck-Institut f\"{u}r Kernphysik, Postfach 103980, D-69029 Heidelberg, Germany.}

\emailAdd{xunjie.xu@gmail.com, hjhe@tsinghua.edu.cn, werner.rodejohann@mpi-hd.mpg.de}

\abstract{
\\
The recent IceCube observation of ultra-high-energy astrophysical neutrinos has begun the era
of neutrino astronomy. In this work, using the unitarity of leptonic mixing matrix, we derive
nontrivial unitarity constraints on the flavor composition of astrophysical neutrinos detected
by IceCube. Applying leptonic unitarity triangles, we deduce these unitarity bounds
from geometrical conditions, such as triangular inequalities. These new bounds generally
hold for three flavor neutrinos, and are independent of any experimental input or the pattern
of leptonic mixing. We apply our unitarity bounds to derive general constraints on
the flavor compositions for three types of astrophysical neutrino sources
(and their general mixture), and compare them with the IceCube measurements.
Furthermore, we prove that for any sources without $\,\nu_\tau^{}$ neutrinos, a detected
$\,\nu_\mu^{}$ flux ratio $\,<1/4\,$ will require the initial flavor composition with
more $\,\nu_e^{}$ neutrinos than $\,\nu_\mu^{}$ neutrinos.
}

\keywords{Neutrino properties, Ultra high energy photons and neutrinos, Neutrino theory
\\[3mm]
JCAP (2014) Final Version, [arXiv:1407.3736].
}

\arxivnumber{1407.3736}

\begin{document}

\maketitle

\setlength{\baselineskip}{18pt}

\setcounter{page}{2}
\vspace*{10mm}
\section{Introduction}
\label{sec:1}
\vspace*{2mm}

With the recent IceCube observation\,\cite{IC0,IC1} of ultra-high-energy astrophysical neutrinos,
the era of neutrino astronomy has finally begun.
The IceCube collaboration has detected a flux of ultra-high-energy cosmic neutrinos
(TeV$-$PeV),  which have $5.7\sigma$ significance above the atmospheric
neutrino backgrounds\,\cite{IC1} and thus point to their extraterrestrial origin.
Contrary to charged particles which would deflect in magnetic fields in space,
such astrophysical neutrinos are expected to point straight back to their sources.
The potential impact of understanding these neutrinos ranges from acceleration mechanisms
of cosmic rays to fundamental particle physics \cite{Winter:2012xq}.
Studying the flavor composition of astrophysical neutrinos
provides an invaluable tool for exploring these issues.
The developments of neutrino telescopes (such as IceCube and alike) \cite{IC0,IC1,NTmore}
in recent years have stimulated extensive studies\,\cite{FR,ratio} on the flavor ratios.
Given these dedicated studies, it is desirable to find general constraints
on the cosmic neutrino flavor compositions.

In this work, we will derive such general constraints by imposing the unitarity of leptonic
mixing matrix \cite{PMNS}, because the leptonic mixing modifies the neutrino flavor ratios
during their trip from source to detector.  The general bounds we obtain do not depend on
the neutrino mixing parameters or any experimental input.
Especially, we will use leptonic unitarity triangles (LUTs) \cite{He:2013rba}\cite{LUTx}
as geometrical means to derive such universal constraints, which turn out to be highly
nontrivial. The unitarity bounds are important, because any violation of these bounds
would call for new physics, such as active-sterile neutrino mixing, neutrino decays,
pseudo-Dirac neutrinos, or other exotic effects \cite{exotic}.

We will then apply our general unitarity bounds to the commonly considered sources of
ultra-high-energy astrophysical neutrinos,
including Pion Sources, Muon-Damped Sources, and Neutron Beam Sources.
We compare these bounds with the IceCube measurement\,\cite{IC1}
and the current global fit of neutrino oscillations\,\cite{fit1,fit2}.
Our unitarity bounds can put general constraints on the emerging flavor ratios
from the IceCube data, independent of specific pattern of leptonic mixing.
Furthermore, we will prove that for any astrophysical sources without $\,\nu_\tau^{}$ neutrinos,
if the detected $\,\nu_\mu^{}$ neutrinos have a flux ratio $\,T< 1/4\,$,\,
then the source must generate more $\,\nu_e^{}$  neutrinos than $\,\nu_\mu^{}$  neutrinos.
These results demonstrate the importance of our general unitarity constraints.
In passing, aspects of the IceCube events were also discussed recently in \cite{ICfit-xi,ICadd}.

The paper is organized as follows.
In Section\,\ref{sec:2}, we will connect the neutrino flavor ratios to the
geometrical parameters of the LUTs. Then, we will use our geometrical formulation to analyze
the general unitarity constraints on the flavor transition probabilities in Section\,\ref{sec:3}.
We apply these constraints to derive nontrivial bounds on the flavor ratios
for typical astrophysical neutrino sources, and compare them with the IceCube data in Section\,\ref{sec:4}.
Finally, we conclude in Section\,\ref{sec:5}.

\vspace*{2mm}
\section{Connecting Astrophysical Neutrinos to Leptonic Unitarity Triangle}
\label{sec:2}
\vspace*{2mm}

The leptonic mixing in charged currents is described by the
$3\!\times\!3$ unitary matrix $U$
of Pontecorvo-Maki-Nakagawa-Sakata (PMNS) \cite{PMNS}.
The orthogonality between the rows (columns) of $\,U\,$ forms the LUTs.
Following the conventions of our recent study\,\cite{He:2013rba},
we define the lengths of the three sides of the LUTs,
\beqa
(a,\,b,\,c) \,\equiv\,
\(|U_{\ell 1}^{}U_{\ell'1}^{}|,\, |U_{\ell 2}^{}U_{\ell'2}^{}|,\, |U_{\ell 3}^{}U_{\ell'3}^{}|\) ,
~~~~
\label{eq:abc}
\eeqa
where the subscripts $\,\ell\,$ and $\,\ell'\,$ stand for the three flavors ($\,\ell\neq\ell'\,$).\,
For each length parameter among $\,(a,\,b,\,c)$,\, we have suppressed the subscripts
$\,\ell\ell'\,$ for simplicity.
The flavor transition probability for astrophysical neutrinos
$\,\nu_{\ell}^{}\to\nu_{\ell'}^{}\,$  is given by
\beqa
P_{\ell\rightarrow\ell'}^{} \,=\,
\sum_{j}^{}|U_{\ell j}^{}U_{\ell'j}^{}|^{2} \,,
\label{eq:P}
\eeqa
which does not contain the oscillation terms since such terms
are simply averaged out due to the very large $L/E$ of astrophysical neutrinos.
From this, we can further express the transition probability
(\ref{eq:P}) in terms of the LUT parameters,
\beqa
P_{\ell\rightarrow \ell'}^{} \,=\, a^{2}+b^{2}+c^{2} \,.
\label{eq:P-abc}
\eeqa
For $\,\ell\neq\ell'$,\, we can classify the flavor appearance probability
$\,P_{\ell\rightarrow \ell'}^{}\,$ into three cases,
\begin{eqnarray}
X & \,=\, & a_{\mu\tau}^{2}+b_{\mu\tau}^{2}+c_{\mu\tau}^{2} \,,
\nn \\[1mm]
Y & \,=\, & a_{\tau e}^{2}+b_{\tau e}^{2}+c_{\tau e}^{2} \,,
\label{eq:0508-1}
\\[1mm]
Z & \,=\, & a_{e\mu}^{2}+b_{e\mu}^{2}+c_{e\mu}^{2} \,.
\nn
\end{eqnarray}
Hence, we can rewrite (\ref{eq:P}) in a matrix form,
\beqa
{\mathbb P} \,=\left(\!
\begin{array}{ccc}
1\!-\!Y\!-\!Z & Z & Y
\\[1mm]
Z & 1\!-\!X\!-\!Z & X
\\[1mm]
Y & X & 1\!-\!X\!-\!Y
\end{array} \!\!\right) \!, ~~~~
\label{eq:0421-1}
\eeqa
where the diagonal elements correspond to survival probability,
\beqa
P_{\ell\rightarrow \ell}^{} \,=\,
1- \sum_{\ell'(\neq \ell)}^{} P_{\ell\rightarrow \ell'}^{} \,,
\label{eq:P-dis}
\eeqa
because the full transition probability equals one.
For an initial flux from a remote astrophysical neutrino source,
let us denote its initial flavor compositions as
$\,(\Phi_{e0}^{},^{}\, \Phi_{\mu 0}^{},\, \Phi_{\tau 0}^{})$.\,
Thus, the detected neutrino flux (after traveling an astronomical distance)
can be computed in the matrix form,
\beqa
(\Phi_{e}^{},\Phi_{\mu}^{},\Phi_{\tau}^{})^{T} \,\propto\,\,
\mathbb{P}\,(\Phi_{e0}^{},\Phi_{\mu0}^{},\Phi_{\tau0}^{})^{T} .
\label{eq:0602-1}
\eeqa

For neutrino telescopes such as IceCube, the high energy muon neutrinos are
in principle distinctive from $\,\nu_{e}^{}\,$ and $\,\nu_{\tau}^{}\,$ signals
as they produce clear muon tracks in the detector.
Hence, the flavor ratio $\,\Phi_{\mu}^{}/\Phit\,$ is a good observable
for these experiments \cite{IC0,IC1,NTmore},
where $\,\Phit =\Phi_{e}^{}+\Phi_{\mu}^{}+\Phi_{\tau}^{}$\,.\,
The other possibly measurable ratio is
$\,\Phi_{e}^{}/\Phit\,$ if the electron neutrino signals can be recognized
in the near future. (The flavor ratio $\,\Phi_{\tau}^{}/\Phit\,$ for tau neutrinos
can be deduced from the other two ratios.)
The $\,\nu_{\mu}^{}\,$ and $\,\nu_{e}^{}\,$ flavor ratios are conventionally defined as
\beqa
T = \frac{\Phi_{\mu}^{}}{\,\Phit\,} \,,
&~~~~&
S = \frac{\Phi_{e}^{}}{\,\Phit\,} \,,
\label{eq:0423-3}
\eeqa
and thus $\,{\Phi_{\tau}^{}}/{\Phit} = 1-T-S\,$.\,
In the literature, sometimes another flavor ratio
$\,R\equiv\Phi_{e}^{}/\Phi_{\tau}^{}\,$ is introduced to replace $\,S\,$.\,
But the description by $(T,\,S)$ is equivalent to that of $(T,\,R)$
because
\beqa
S \,=\, (1\!-\!T)\frac{R}{\,1\!+\!R\,}\,,
&~~~~&
R \,=\, \frac{S}{\,1\!-\!T\!-\!S\,\,} \,.
\label{eq:0602-2}
\eeqa

Inspecting the formulas \eqref{eq:0508-1}, we find that under the exchange
$\,\nu_e^{}\leftrightarrow\nu_\mu^{}\,$,\,
the transition probabilities $(X,\,Y,\,Z)$ change as follows:
$\,X\leftrightarrow Y\,$ and $\,Z\leftrightarrow Z\,$.\,
This also corresponds to the exchanges of the first and second rows (columns)
of the matrix $\,\mathbb{P}\,$ in Eq.\,\eqref{eq:0421-1}.
With Eqs.\,\eqref{eq:0602-1} and \eqref{eq:0423-3}, we further infer
$\,T\leftrightarrow S\,$ under the same exchange of
$\,\nu_e^{}\leftrightarrow\nu_\mu^{}\,$.\,

\vspace*{2mm}

Typically, let us consider three types of commonly studied neutrino sources, i.e.,
Pion Sources ($\pi$S), Muon-Damped Sources ($\mu$DS), and Neutron Beam Sources ($n$BS).
\begin{itemize}

\item
The $\pi$S sources produce neutrinos from pion decays,
$\,\pi\rightarrow\mu+\nu_{\mu}^{}\rightarrow e +\nu_{e}^{}+2\nu_{\mu}^{}$,\,
where we do not distinguish the notations between particles and anti-particles for
simplicity. Hence, the initial flavor composition is $\,(1:2:0)$\,.\,
From Eq.\,(\ref{eq:0602-1}), the $\,\nu_{\mu}^{}\,$ and $\,\nu_{e}^{}\,$ flux ratios
in this case are given by
\begin{equation}
T\,= \fr{1}{3}\! \(2 \!-\! 2X \!-\! Z\), ~~~
S\,= \fr{1}{3}\! \(1 \!-\! Y \!+\! Z\).
\label{eq:0423}
\end{equation}

\item
The $\mu$DS sources produce muon neutrinos in
$\,\pi\rightarrow\mu+\nu_{\mu}\,$,\,
where the damped muons lose energy so that the neutrino flux produced from their
decays is depleted at energies of interest.
Hence, the initial flavor composition is $(0:1:0)$.\,
From Eq.\,(\ref{eq:0602-1}), we have the
$\,\nu_{\mu}^{}\,$ and $\,\nu_{e}^{}\,$ flux ratios,
\beqa
T \,= 1\!-\!X\!-\!Z \,, ~~~~
S \,= Z \,.
\label{eq:0423-1}
\eeqa

\item
The Neutron Beam Sources ($n$BS) produce electron neutrinos in beta
decay of neutrons. Thus, its initial flavor composition is $(1:0:0)$.\,
From Eq.\,(\ref{eq:0602-1}), the $\,\nu_{\mu}^{}\,$ and $\,\nu_{e}^{}\,$
flux ratios are given by
\beqa
T \,= Z\,, ~~~~
S \,= 1\!-\!Z\!-\!Y \,.
\label{eq:0423-2}
\eeqa

\end{itemize}

For the current experiments, which source the detected high-energy astrophysical neutrinos
originate from is uncertain. Nevertheless, we note that if all three
types of sources are involved, the initial neutrino flux would contain no
$\nu_{\tau}^{}$ neutrinos. Let us consider a general source with mixture\,\cite{FR} from
all three types of sources above. In this case, the initial flavor composition can be written as,
$\,\(\eta :1\!-\!\eta :0\)$,\, with the parameter $\,\eta \in [0,\,1]$\,.\,
Hence, in the general case, we have $\,T\,$ and $\,S\,$ flux ratios
depending on $\,\eta\,$,
\beqs
\label{eq:TS-XYZ-eta}
\begin{eqnarray}
T & \,=\, & \eta\, Z+(1\!-\!\eta )(1\!-\!X\!-\!Z) \,,
\label{eq:0602-3}
\\[1.5mm]
S & \,=\, & \eta \(1\!-\!Z\!-\!Y\)+(1\!-\eta )Z \,.
\label{eq:0602-4}
\end{eqnarray}
\eeqs

\vspace*{2mm}
\section{Unitarity Constraints on Flavor Transitions of Astrophysical Neutrinos}
\label{sec:3}
\vspace*{1mm}

The leptonic mixing matrix of PMNS \cite{PMNS} is unitary,
$\,UU^{\dagger}=I\,$,\,
which imposes two kinds of constraints on the row vectors
$(U_{\ell 1}^{},\,U_{\ell 2}^{},\, U_{\ell 3}^{})$.\,
These include, (i) the normalization conditions,
\beqa
\label{eq:U-i}
|U_{\ell 1}^{}|^{2}+|U_{\ell 2}^{}|^{2}+|U_{\ell 3}^{}|^{2} = 1 \,,
\eeqa
and (ii) the orthogonal conditions,
\beqa
\label{eq:U-ii}
U_{\ell 1}^{*}U_{\ell'1}^{}\!+U_{\ell2}^{*}U_{\ell'2}^{}\!+U_{\ell 3}^{*}U_{\ell' 3}^{}=0\,,
~~~(\ell\neq\ell')\,.~~
\eeqa

The second constraint \eqref{eq:U-ii} implies the closure of the corresponding unitarity
triangle, since the three complex numbers can be represented by three vectors
in the complex plane and the zero sum makes them form a closed triangle.
In terms of the lengths of three sides $\,(a,\,b,\,c)$,\,
the closure imposes nontrivial triangular inequalities, stating that
the sum of the lengths of any two sides is larger than the remaining side,
\beqa
a+b\geqq c\,, ~~~ a+c\geqq b\,, ~~~ b+c\geqq a\,,
\label{eq:0507}
\eeqa
where the equality sign corresponds to the collapse of the triangle into a line.
Another equivalent statement is that the difference between
the lengths of any two sides is smaller than the remaining side,
because  $\,a-b\leqq c$\, is just $\,b+c\geqq a$\,,\, and so on.
Hence, Eq.\,(\ref{eq:0507}) is sufficient to describe the
triangular closure constraints.

The geometrical meaning of the first constraint \eqref{eq:U-i} does not appear
so obvious, but in fact it restricts the length scale of the three sides.
Let us define the notations,
$\,(a_{1}^{},\,b_{1}^{},\,c_{1}^{})\equiv(|U_{\ell 1}^{}|,\,|U_{\ell 2}^{}|,\,|U_{\ell 3}^{}|)$
and
$\,(a_{2}^{},\,b_{2}^{},\,c_{2}^{})\equiv(|U_{\ell'1}^{}|,\,|U_{\ell'2}^{}|,\,|U_{\ell'3}^{}|)$.\,
Thus, we can express the three sides,
$(a,\,b,\,c)=(a_{1}^{}a_{2}^{},\,b_{1}^{}b_{2}^{},\,c_1^{}c_2^{})$.\,
Using the Cauchy\textendash{}Schwarz inequality\footnote{
The Cauchy\textendash{}Schwarz inequality is
well-konwn in mathematics, which states,
$\,(x_{1}^{}y_{1}^{}+x_{2}^{}y_{2}^{}+ \cdots +x_{n}^{}y_{n}^{})^{2}
\leqq (x_{1}^{2}+x_{2}^{2}+\cdots+x_{n}^{2})(y_{1}^{2}+y_{2}^{2}+\cdots+y_{n}^{2})$\,.\,
Another key inequality we use in this work is that the arithmetical mean
is smaller than the corresponding quadratic mean,
$\,(x_{1}^{}+x_{2}^{}+\cdots +x_{n}^{})/n \leqq
   \sqrt{(x_{1}^{2}+x_{2}^{2}+ \cdots +x_{n}^{2})/n\,}\,$.\,
For more detail, see for instance,
G.\ H.\ Hardy, J.\ E.\ Littlewood, and G.\ Polya,
\emph{Inequalities,} 1952, Cambridge University Press.},\,
we deduce
\beqa
 (a_{1}a_{2}+b_{1}b_{2}+c_{1}c_{2})^{2}
 \,\leqq\, (a_{1}^{2}+b_{1}^{2}+c_{1}^{2})(a_{2}^{2}+b_{2}^{2}+c_{2}^{2})
 \,\leqq\,  1 \,.
\label{eq:0421-2}
\eeqa

From this, we deduce that the perimeter of the triangle
(the sum of its three sides) cannot exceed one,
\beqa
a+b+c \,\leqq\, 1 \,.
\label{eq:0507-1}
\eeqa
With this, we can further derive an upper bound on the Jarlskog invariant
$\,J = \rm{ Im }\{U_{\ell' j}^{} U_{\ell j}^\ast U_{\ell k}^{} U_{\ell' k}^\ast\}$\,,\,
with $\,\ell \neq \ell'$\, and $j \neq k$ \cite{J}, fully from geometry.
The Euclidean geometry tells us that
a shape with fixed perimeter reaches the maximal area when it is a circle,
and for a triangle with fixed perimeter, its maximal area is realized
when it is an equilateral triangle, with $\,a=b=c\,$ and the corresponding
area $\,S_{\max}^{} = \sqrt{3}a^{2}/4$.\,
(Intuitively, the equilateral triangle in some sense looks like a circle more than
any other triangles.)
Since the Jarlskog invariant equals twice the area of the LUT, the
maximum $\,|J|\,$ is given by the equilateral unitarity triangle,
$\,|J|_{\max}^{} = \sqrt{3}a^{2}/2$\, with $\,a=1/3\,$.\,
Hence, without using any parameter from the conventional PMNS matrix,
we can derive the general geometrical upper bound on $\,|J|\,$,
\beqa
|J| \,\leqq\, \frac{1}{\,6\sqrt{3}\,} \,.
\label{eq:0507-2}
\eeqa
\begin{figure}
\centering
\includegraphics[width=8.1cm]{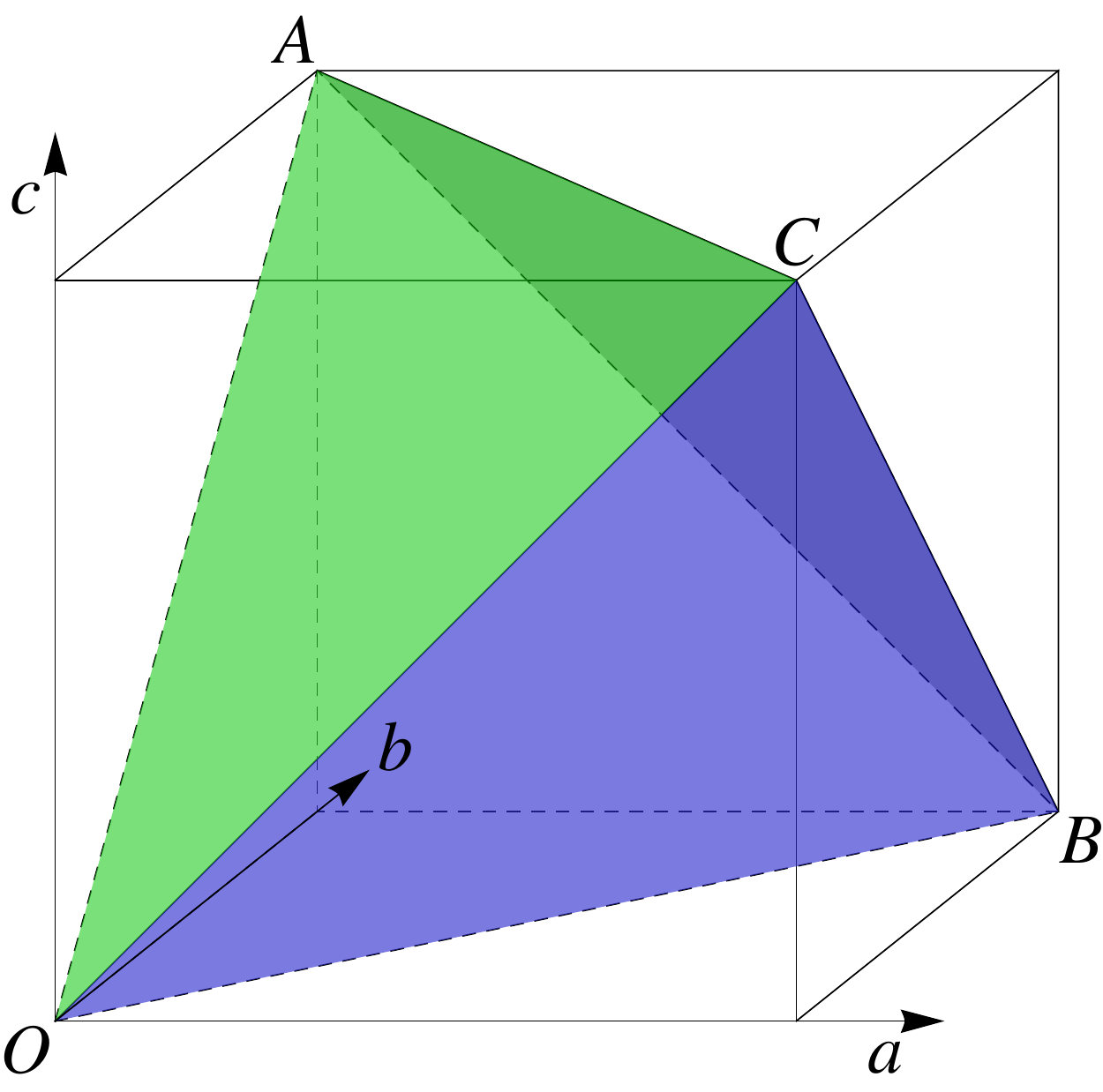}
\caption{Unitarity constraints on the lengths of sides for any
leptonic unitarity triangle, $(a,\,b,\,c)$.\,  Each side of the cube
has length equal $\,1/2$,\, and thus $\,(a,\,b,\,c)$\, cannot be
larger than $\,1/2$.\, The equations of planes $ABC$, $OAC$, $OBC$
and $OAB$ are $\,a+b+c=1$,\, $\,a+b=c$,\, $b+c=a$,\, and $\,c+a=b$,\,
respectively. The unitarity requires $\,(a,\,b,\,c)$\, to be a point
inside the tetrahedron $OABC$.}
\label{fig:1}
\vspace*{2mm}
\end{figure}

Even though the condition (\ref{eq:0507-1}) appears quite different
from (\ref{eq:0507}), geometrically they are very similar
as Fig.\,\ref{fig:1} illustrates.   Each side of the cube in Fig.\,\ref{fig:1} has
length equal $\,1/2\,$.\,  Hence, the equation of the plane $ABC$ is, $\,a+b+c=1$\,.\,
The inequality (\ref{eq:0507-1}) is derived from the normalization condition
\eqref{eq:U-i}, and it requires that the allowed
region should be on one side of the plane $ABC$.
The other planes of the tetrahedron are planes $OAC$, $OBC$ and $OAB$,
corresponding to $\,a+b=c\,$,\, $b+c=a\,$ and $\,c+a=b\,$,\, respectively.
These planes make a tetrahedron with each side of length $\,1/\!\sqrt{2}\,$.\,
The inequalities (\ref{eq:0507-1}) and (\ref{eq:0507}) only require that
$\,(a,\,b,\,c)$\, is a point inside the tetrahedron.
Thus, we can immediately infer the upper bound on the length of each side
for any LUT,
\beqa
a,b,c \,\leqq\, \frac{1}{2} \,.
\label{eq:0507-3}
\eeqa

Next, we would ask: what are the unitarity bounds on the averaged transition probabilities
$\,(X,\,Y,\,Z)$\, defined in Eq.\,(\ref{eq:0508-1})?
Here, we can deduce and visualize the bounds geometrically.
Consider a sphere with its center at $\,(0,0,0)$.\,
The sphere retains some of the allowed points $\,(a,\,b,\,c)$\, on it,
and has intersections with the tetrahedron. It should have a radius no
larger than $\,1/\!\sqrt{2}$\,.\, Hence, we deduce
\beqa
a^{2}+b^{2}+c^{2} \,\leqq\, \frac{1}{2} \,.
\label{eq:0508}
\eeqa
Using Eq.\,(\ref{eq:0508-1}), we infer the nontrivial upper bound,
\beqa
X,Y,Z ~\leqq~ \frac{1}{2} \,.
\label{eq:0508-2}
\eeqa
We stress that we derived these constraints only from the unitarity of the PMNS matrix,
without any experimental input.   This means that for astrophysical neutrinos
(or any neutrinos traveling with a large enough $L/E$\,),
the flavor appearance probability for any two flavors ($\ell\to\ell'$)
cannot exceed $\,1/2$\,,
\begin{equation}
P_{\ell\rightarrow\ell'}^{} \,\leqq\, \frac{1}{2} \,.
\label{eq:0508-3}
\end{equation}

Another nontrivial result we will prove is that the survival probability
is bounded from below, always no smaller than $\,1/3$\,,
\begin{equation}
P_{\ell\rightarrow\ell}^{} \,\geqq\, \frac{1}{3} \,.
\label{eq:0508-4}
\end{equation}
The survival probability $\,P_{\ell\rightarrow\ell}^{}\,$ is just the diagonal
elements of the matrix (\ref{eq:0421-1}).  To prove (\ref{eq:0508-4}), we first choose
$\,\ell=e\,$ for definiteness, $\,P_{e\rightarrow e}^{} =1-Y-Z$\,.\,
Note that
$\,Y+Z =
   (a_{\tau e}^{2}\!+b_{\tau e}^{2}\!+c_{\tau e}^{2})
  +(a_{e\mu}^{2}\!+b_{e\mu}^{2}\!+c_{e\mu}^{2})$,\,
where the terms
$\,a_{\tau e}^{2}\!+a_{e\mu}^{2}$,\, for instance, can be written as
\begin{eqnarray}
a_{\tau e}^{2}\!+a_{e\mu}^{2}
& \,=\, & |U_{e1}^{}|^{2}\(|U_{\mu 1}^{}|^{2}\!+|U_{\tau 1}^{}|^{2}\)
\nonumber \\[2mm]
& \,=\, & |U_{e1}^{}|^{2}\(1-|U_{e1}^{}|^{2}\) .
\label{eq:0509-1}
\end{eqnarray}
We can derive similar formulas for $\,b_{\tau e}^{2}+b_{e\mu}^{2}\,$
and $\,c_{\tau e}^{2}+c_{e\mu}^{2}\,$.\,
With these, we arrive at
\begin{eqnarray}
Y+Z & \,=\, & \sum_{j=1}^{3}|U_{ej}^{}|^{2}(1-|U_{ej}^{}|^{2})
 \,=\, 1-\sum_{j=1}^{3}|U_{ej}^{}|^{4} ~~~~
\nonumber \\
& \,\leqq\, & 1-\frac{1}{3}\Big(\sum_{j=1}^{3}|U_{ej}|^{2}_{}\Big)^2
  \,\leqq\, \frac{2}{3} \,.
\label{eq:0509}
\end{eqnarray}
This leads to $\,P_{e\rightarrow e}^{}\geqq \frac{1}{3}$\,.\,
The first inequality in the second line of  Eq.\,(\ref{eq:0509}) is based on the fact
that the arithmetic mean of several real numbers is always smaller than their
quadratic mean (cf.\ footnote-1).
Likewise, we can prove that $\,X+Y\,$ and $\,Z+X\,$
obey the same inequality,
\beqa
X\!+\!Y,\,Y\!+\!Z,\,Z\!+\!X \,\leqq\, \frac{2}{3} \,.
\hspace*{8mm}
\label{eq:0509-2}
\eeqa
With these, we complete the proof of the lower bound (\ref{eq:0508-4}) on the survival probability.

Furthermore, we will prove the following nontrivial inequalities,
\beqs
\label{eq:bound-Y+2Z}
\begin{eqnarray}
Y\!+\!2Z,\, Z\!+\!2X,\, X\!+\!2Y & \,\leqq\, & \frac{25}{24} \,,
\label{eq:0515}\\[2mm]
2Y\!+\!Z,\, 2Z\!+\!X,\, 2X\!+\!Y & \,\leqq\, & \frac{25}{24} \,.
\label{eq:0515-1}
\end{eqnarray}
\eeqs
We present the proof as follows.
Without losing generality, we take $\,Y\!+2Z\,$ for instance.
Let us inspect the difference,
\begin{eqnarray}
G
& \,\equiv\, & (Y\!+2Z)-1
\nn\\[1.5mm]
& = & \sum_{j}|U_{ej}|^{2}\(2|U_{\mu j}|^{2}+|U_{\tau j}|^{2}-1\)
\nn \\
& = & \sum_{j}|U_{ej}|^{2}\(|U_{\mu j}|^{2}-|U_{ej}|^{2}\) .
\label{eq:0520-1}
\end{eqnarray}

Our proof will be achieved so long as we demonstrate the maximum value,
$\,G_{\max}^{}=\frac{1}{24}\,$.\,  Since $\,G\,$ only depends
on the first two rows of the PMNS matrix, we can generally write the squared elements
in a matrix form,
\begin{equation}
|U_{\ell\ell'}^{}|^{2} \,\equiv \left(\!\begin{array}{ccc}
x~ & ~y~ & ~1\!-\!x\!-\!y
\\[1mm]
z~ & w & ~1\!-\!z\!-\!w
\\[1mm]
\times~ & \times & ~\times
\end{array}\!\right) \!,
\label{eq:0520}
\end{equation}
where ``$\times$'' denote elements of no interest here.
The quantity $\,G\,$ is a function of $\,(x,\,y,\,z,\,w)$,\, which may be regarded as
equivalents to the four independent parameters of the PMNS matrix.
Using the notation \eqref{eq:0520}, we can rewrite the function $\,G$\,,
\begin{eqnarray}
G & \,=\, & xz+yw+(1\!-\!x\!-\!y)(1\!-\!z\!-\!w)
 -\left[ x^{2}+y^{2}+(1\!-\!x\!-\!y)^{2}\right] \!.
\label{eq:0520-2}
\end{eqnarray}
If we overlook the boundary of parameter space, we would naively seek
the maximum by solving
$\,\partial_{x}G=\partial_{y}G=\partial_{z}G=\partial_{w}G=0$\,.\,
This gives a unique solution, $\,x=y=z=w=\frac{1}{3}$\,,\, which results in $\,G=0\,$.\,
But, as can be readily checked,
this solution is actually a saddle point, rather than the maximum.
This implies that the maximum of $\,G\,$ should be on the boundary, since
this saddle point is the only place where the first derivatives of
$\,G\,$ vanish.  Hence, we will inspect the maximum of $\,G\,$ on the boundary
of parameter space.

The relevant parameter space is where $\,(x,\,y,\,z,\,w)\,$ satisfies
(i) $\,x,\,y,\,z,\,w\geqq 0\,$ and $\,x+y,\,z+w\leqq 1\,$,\,
and (ii) the triangle inequalities $\,a+b\geqq c,\, a+c\geqq b,\, b+c\geqq a\,$,\,
where $\,a=\sqrt{xz}$\,,\, $b=\sqrt{yw}$\,,\, and $\,c=\sqrt{(1\!-\!x\!-\!y)(1\!-\!z\!-\!w)}$\,.\,
Any \,$(x,\,y,\,z,\,w)$\, satisfying these two conditions can realize a unitary PMNS matrix.
When we are on the boundary of the condition (i), then the first two rows of $(\ref{eq:0520})$
must contain one zero element. We will prove that only when the second row has a zero element,
$G$ realizes its global maximum.
In this case, without losing generality, we set the
third element of this row be zero, i.e., $\,1-z-w=0\,$,\, then we can resolve
$\,\partial_{x,y,z}^{}G|_{w=1-z}=0$\,.\,
We find the solution, $\,(x,\,y,\,z)=(\frac{5}{12},\,\frac{5}{12},\,\frac{1}{2})\,$
and $\,w=1-z=\frac{1}{2}\,$.\,
This gives the maximum,
\beqa
G_{\max} \,=\, \frac{1}{24} \,.
\label{eq:0520-3}
\eeqa

Next, we will prove that the other cases either have no extremum or have the extremum
not as a global maximum. If the maximum of $\,G\,$ is on the boundary of the
condition (i), but with the zero element in the first row of \eqref{eq:0520}, we may set $\,1-x-y=0$\,
without losing generality. In this case, we find that the extremum equation
$\,\partial_{x,z,w}^{}G|_{x=1-y}^{}=0\,$
has no solution by direct calculation.

\begin{figure}[t]
\centering
\includegraphics[width=8.1cm]{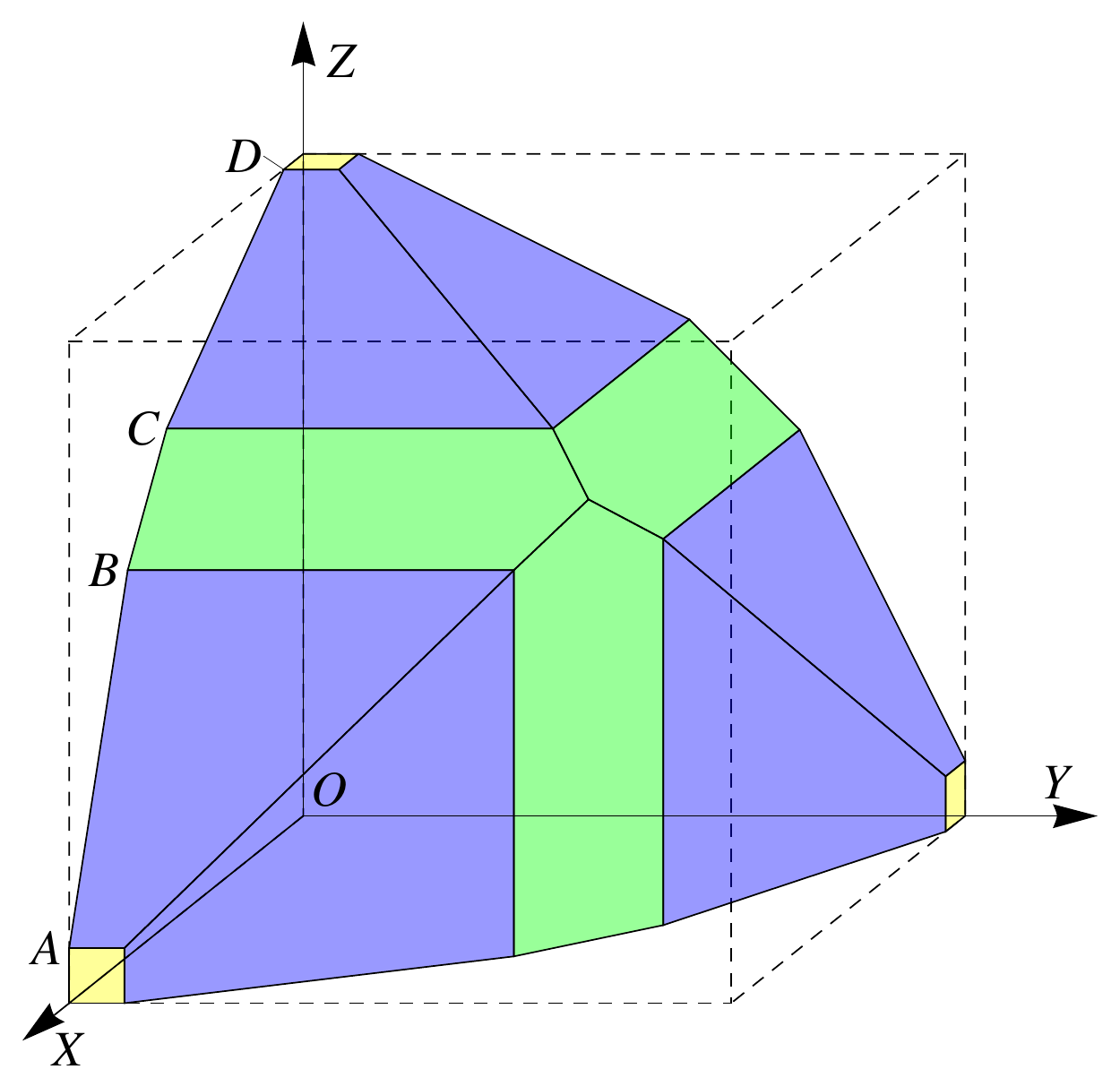}
\caption{Unitarity constraints on the averaged neutrino transition
probabilities $(X,\,Y,\,Z)$.\, Each side of the cube has length equal to $\,\frac{1}{2}\,$.\,
The conditions (\ref{eq:0508-2}) confine the $(X,\,Y,\,Z)$ space into the cube.
The equations for the green planes are
$\,(X\!+\!Y,\, Y\!+\!Z,\, Z\!+\!X)=\frac{2}{3}$\,,\, while for the blue planes they are
$\,(2X\!+\!Y,\,2X\!+\!Z,\,2Y\!+\!Z,\,2Y\!+\!X,\,2Z\!+\!X,\,2Z\!+\!Y)=\frac{25}{24}\,$.\,
The unitarity of the PMNS matrix requires $\,(X,\,Y,\,Z)$\, to be a point
inside the region bounded by the colored surfaces.}
\label{fig:2}
\vspace*{2mm}
\end{figure}

If the maximum is instead on the boundary of the condition (ii),
we have one of the triangle inequalities saturated. Since it is not on the boundary
of the condition (i), all the elements of the two rows are non-zero,
which means that $(a,\,b,\,c)$ are all non-zero.
Hence, only one of the triangle inequalities can be saturated. Without losing
generality, we consider the situation $\,a+b=c\,$.\,
This is a hypersurface, $\,F(x,y,z,w)=0\,$,\, in the parameter space where
\beqa
\hspace*{-3mm}
F \,=\, (a+b)^2-c^2
\,=\, xz \!+\! yw \!+\! 2\sqrt{xyzw} \!-\! (1\!-\!x\!-\!y)\(1\!-\!z\!-\!w\).~~~~
\label{eq:0520-4}
\eeqa
The extremum point can be found by the method of Lagrange multipliers.
This is to solve $\,\partial_{x,y,z,w}^{}(G+\lambda F)=0\,$ and $\,F=0\,$
as five equations for $\,(x,\,y,\,z,\,w,\,\lambda)$\,.\,
The function $G$ constrained on the hypersurface reaches an extremum with,
$\,(x,\,y,\,z,\,w)=(\frac{7}{24},\,\frac{7}{24},\,\frac{5}{24},\,\frac{5}{24})$
and $\,\lambda=\frac{1}{8}$.\,
At this point, we find $\,G=\frac{1}{48}$\,,\, which is less than
\eqref{eq:0520-3}. Hence, it is not the global maximum.
This completes our proof of \eqref{eq:0520-3} and thus the bounds
\eqref{eq:bound-Y+2Z}.

The inequalities (\ref{eq:0508-2}), (\ref{eq:0509-2}) and \eqref{eq:bound-Y+2Z}
impose nontrivial unitarity bounds on the transition probabilities
$(X,\,Y,\,Z)$.\,  We present these bounds in Fig.\,\ref{fig:2}, where
the allowed region is surrounded by the colored surfaces.
First, the conditions of (\ref{eq:0508-2}) restrict $\,(X,\,Y,\,Z)$\,
into a cube (yellow color) with each side length equal to $\,\frac{1}{2}$\,.\,
Second, the inequalities of (\ref{eq:0509-2}) constrain the region through the
three green planes, described by the equations
$\,(X\!+\!Y,\, Y\!+\!Z,\, Z\!+\!X)=\frac{2}{3}$\,.\,
Finally, Eq.\,\eqref{eq:bound-Y+2Z} further bounds the allowed
region through the six blue planes, dictated by the equations
$\,(2X\!+\!Y,\,2X\!+\!Z,\,2Y\!+\!Z,\,2Y\!+\!X,\,2Z\!+\!X,\,2Z\!+\!Y)=\frac{25}{24}\,$.\,

\vspace*{3mm}
\section{Unitarity Constraints on Flavor Ratios of Astrophysical Neutrinos}
\label{sec:4}
\vspace*{1mm}

As mentioned earlier, the astrophysical neutrinos may originate from different sources.
The commonly considered neutrino sources include Pion Sources
($\pi$S), Muon-Damped Sources ($\mu$DS), and Neutron Beam Sources ($n$BS).
In this section, we will apply the general unitarity bounds
(\ref{eq:0508-2}), (\ref{eq:0509-2}) and \eqref{eq:bound-Y+2Z}
to derive new constraints
on the flavor ratios for different sources of cosmic neutrinos.
These general constraints are independent of any experimental input or
specific pattern of leptonic mixing.

\vspace*{4mm}
\subsection{Pion Sources with Flavor Ratio (1\,:\,2\,:\,0)}
\vspace*{2mm}

As mentioned in Sec.\,2,
the Pion Sources have the initial neutrino flavor ratio equal (1\,:\,2\,:\,0).\,
Thus, we can deduce the flavor ratios at the detector as in \eqref{eq:0423},
$\,T=\frac{1}{3}(2-2X-Z)$\, and $\,S=\frac{1}{3}(1-Y+Z)\,$.\,

From (\ref{eq:0508-2}) and \eqref{eq:bound-Y+2Z}, we have,
$\,-\frac{1}{2}\leqq -Y\leqq Z-Y\leqq Z\leqq\frac{1}{2}$\, and
$\,0\leqq 2X+Z\leqq \frac{25}{24}$\,.\, Thus, we can deduce
\beqs
\label{eq:120-UB-TS}
\beqa
\fr{23}{72} \,\leqq &~T~& \leqq\, \fr{2}{3} \,,
\label{eq:120-UB-T}   
\\[2mm]
\fr{1}{6} \,\leqq &~S~& \leqq\, \fr{1}{2} \,.
\label{eq:120-UB-S}   
\eeqa
\eeqs

Next, we will analyze the unitarity bounds for
\,$S\!+\!T$,\, $S\!-\!T$,\, $T\!+\!2S$,\, $T\!+\!4S$\, and \,$3S\!-\!T$.\,
We may first compute the combinations of $\,T\,$ and $\,S\,$,\,
%
\beqa
S+T &\,=\,& 1-\fr{1}{3}(2X+Y) \,,
\label{eq:0505-1-1}
\nn\\[1.5mm]
S-T &\,=\,& -\fr{1}{3}+\fr{1}{3}(2X+2Z-Y) \,,
\label{eq:0505-1-1-1}
\nn\\[1.5mm]
T+2S  &\,=\,& \fr{1}{3}(4-2X-2Y+Z) \,,
\label{eq:0505-3-1}
\\[1.5mm]
T+4S &\,=\,& 2-\fr{1}{3}(2X+4Y-3Z) \,,
\label{eq:0515-2}
\nn\\[1.5mm]
3S-T &\,=\,& \fr{1}{3}(1-3Y+2X+4Z) \,.
\nn\label{eq:0517}
\eeqa
%

From the conditions (\ref{eq:0508-2}) and \eqref{eq:bound-Y+2Z}, we infer
the unitarity bounds on $\,S\!+\!T\,$,
\begin{equation}
\fr{47}{72} \,\leqq\, S\!+\!T \,\leqq\, 1 \,.
\label{eq:120-UB-S+T}    
\end{equation}
The flavor ratio difference $\,S\!-\!T$\, in (\ref{eq:0505-1-1-1}) contains
$\,2X+2Z-Y\,$,\, which is larger than $\,-Y\,$ and smaller than $\,2(X+Z)\,$.\,
Thus, from (\ref{eq:0509-2}) and \eqref{eq:bound-Y+2Z} we derive,
\begin{equation}
-\!\fr{1}{2} \,\leqq\, 2X\!+\!2Z\!-\!Y \,\leqq\, \fr{4}{3} \,,
\label{eq:0505-4-1}
\end{equation}
which leads to the bound,
\begin{equation}
-\!\fr{1}{2} \,\leqq\, S\!-\!T \,\leqq\, \fr{1}{9} \,.
\label{eq:120-UB-S-T}    
\end{equation}
We note that $\,T+2S\,$ contains the combination $\,2X\!+\!2Y\!-\!Z$,\,
which subjects to the same bounds as in Eq.\,(\ref{eq:0505-4-1}). 
Hence, we arrive at
\beqa
\fr{8}{9} \,\leqq\, T\!+\!2S \,\leqq\, \fr{3}{2} .
\label{eq:120-UB-T+2S}
\eeqa
The upper and lower bounds on $\,2X+4Y-3Z$\, or $\,2X+4Z-3Y$\,
are $\,\frac{25}{12}\,$ and $\,-\frac{3}{2}\,$,\, respectively, which can be
inferred in a similar way to (\ref{eq:0505-4-1}).
Hence, we can deduce
\begin{equation}
\fr{47}{36} \,\leqq\, T\!+\!4S \,\leqq\, \fr{5}{2} ,
\label{eq:120-UB-T+4S}    
\end{equation}
and
\begin{equation}
-\!\fr{1}{6} \,\leqq\, 3S\!-\!T \,\leqq\, \fr{37}{36} .
\label{eq:120-UB-3S-T}   
\end{equation}

\begin{figure}[t]
\centering
\includegraphics[height=9.5cm,width=9.5cm]{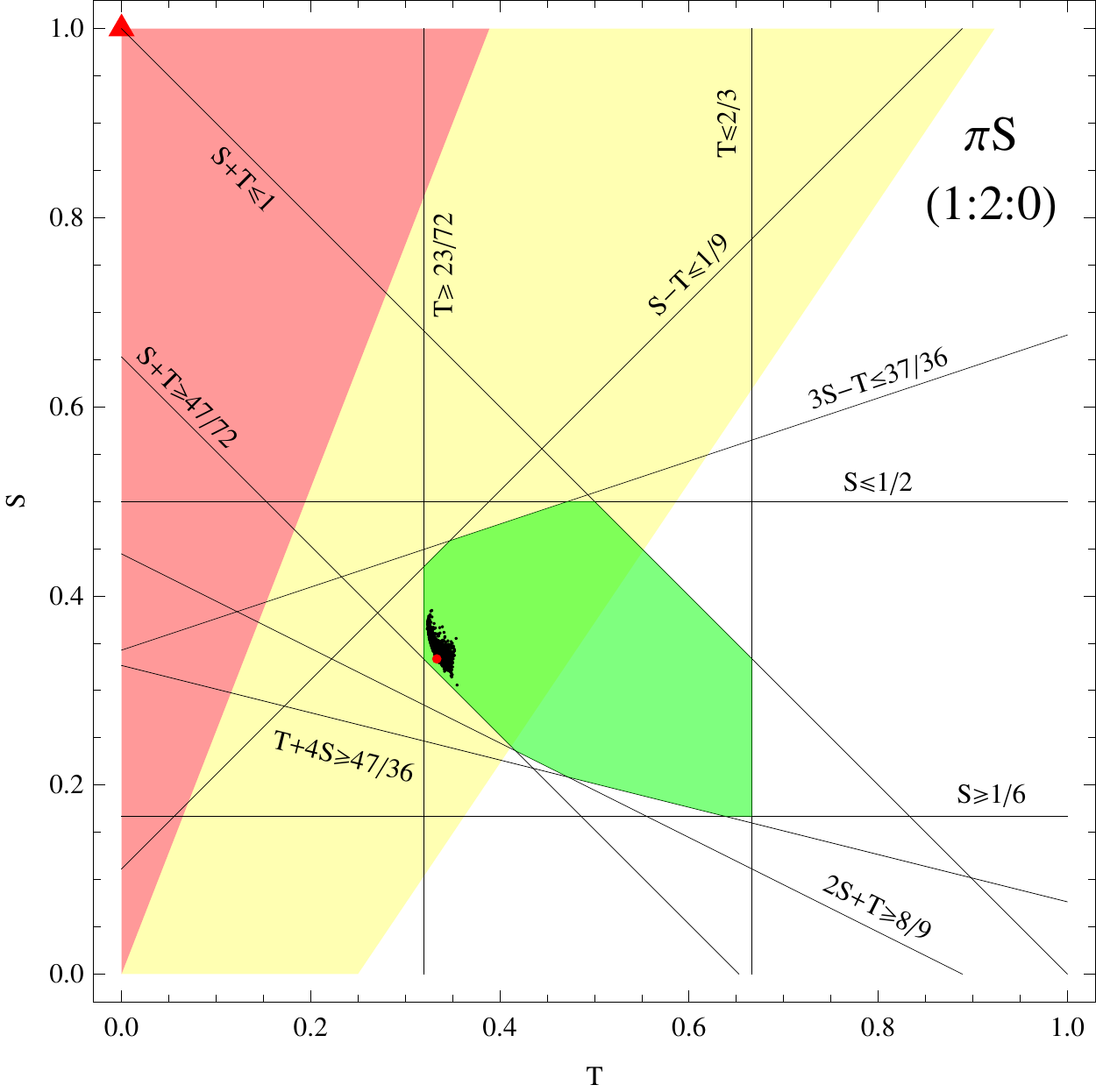}
\vspace*{-2mm}
\caption{Unitarity bounds on the flavor ratios $\,(T,\,S)$,\,
where we take Pion Sources ($\pi$S) with initial neutrino flavor ratio \,(1\,:\,2\,:\,0)\,.\,
The black straight lines represent the general bounds (\,Table\,\ref{tab:1}\,) derived from the unitarity
of the PMNS matrix without experimental input. These bounds are combined to give
the allowed region (green area). The dark spot inside the shaded area is a collection
of 1000 random points given by the results of a neutrino global fit\,\cite{fit1} of the PMNS matrix.
The red point nearby the boundary of the green area depicts $\,(T,\,S)=(\frac{1}{3},\,\frac{1}{3})$,
and corresponds to a flux ratio of \,(1\,:\,1\,:\,1)\, at the detector.
The shaded red (yellow) region denotes the fit\,\cite{ICfit-xi} to the three-year IceCube data
at 68\%\,C.L.\,(95\%\,C.L.), with the red-triangle as the best fit.
}
\label{fig:3}
\vspace*{2mm}
\end{figure}

We summarize the above unitarity bounds \eqref{eq:120-UB-TS}, \eqref{eq:120-UB-S+T} and
\eqref{eq:120-UB-S-T}-\eqref{eq:120-UB-3S-T} in Table\,\ref{tab:1}, for Pion Sources
with the initial flavor ratio \,(1\,:\,2\,:\,0).\,
Combining all these constraints, we identify the allowed region of \,$(T,\,S)$\, in Fig.\,\ref{fig:3},
which lies in the shaded area (green color).
In Fig.\,\ref{fig:3}, we also present the parameter region (black points) allowed
by the current neutrino global fit \cite{fit1},
\begin{eqnarray}
s_{12}^{2} &\,=\,& (3.08\pm 0.17)\!\times\! 10^{-1},
\nn\\[1.5mm]
s_{23}^{2} &=& (4.37\pm 0.28)\!\times\! 10^{-1},
\label{eq:0511}
\\[1.5mm]
s_{13}^{2} &=& (2.34\pm 0.20)\!\times\! 10^{-2},
\nn\\[1.5mm]
\delta_{D}^{} &=& (1.39\pm 0.33)\pi \,.
\nn
\end{eqnarray}
The global fit \eqref{eq:0511} is for the normal mass-ordering. As we have checked,
the global fit for inverted mass-ordering only differs a little,
and does not lead to any visible effect in our numerical analyses.
Thus, it suffices to use the above fit \eqref{eq:0511} for our present study.
Using the global fit \eqref{eq:0511} for the PMNS parameters
$\,(s_{13}^{},\,s_{23}^{},\,s_{12}^{},\,\delta_{D}^{})$\, with Gaussian distributions,
we have generated 1000 random points in Fig.\,\ref{fig:3}.
From this plot, we see that these black points appear nearly as a dark spot
in the small region of the $\,T-S\,$ plane, as required by the current neutrino global fit.

%
\begin{center}
\begin{table}[t]
\caption{Summary of unitarity constraints on the flavor ratios of astrophysical neutrinos.}
\label{tab:1}
\vspace*{3mm}
\begin{tabular}{c|c|c}
\hline\hline
&&
\\[-3.5mm]
~$\nu$~Sources~ & Initial Ratio & Leptonic Unitarity Bounds
\\[-3.5mm]
&&
\tabularnewline
\hline
\hline
&&
\\[-3.5mm]
$\pi$S 
& (1\,:\,2\,:\,0) &
~$\frac{23}{72}\leqq T\leqq\frac{2}{3}$,\, $\frac{1}{6}\leqq S\leqq\frac{1}{2}$,\,
$\frac{47}{72}\leqq T\!+\!S\leqq 1$,\, $-\frac{1}{9}\leqq T\!-\!S\leqq\frac{1}{2}$~
\tabularnewline[0.1cm]
 &  &  $\frac{8}{9}\leqq T\!+\!2S\leqq\frac{3}{2}$,\, $\frac{47}{36}\leqq T\!+\!4S\leqq\frac{5}{2}$,\,
$-\frac{37}{36}\leqq T\!-\!3S\leqq\frac{1}{6}$
\\[-3.5mm]
&&
\tabularnewline[0.1cm]
\hline
&&
\\[-3.5mm]
$\mu$DS  
& (0\,:\,1\,:\,0) & $\frac{1}{3}\leqq T\leqq 1$,\,
$0\leqq S\leqq\frac{1}{2}$,
\tabularnewline[1mm]
 &  &  $\frac{1}{2}\leqq T\!+\!S\leqq 1$,\,
 $-\frac{1}{24}\leqq T\!-\!S\leqq 1$,\, $\frac{23}{24}\leqq 2T\!+\!S\leqq 2$
\\[-3mm]
&&
\tabularnewline
\hline
&&
\\[-3.5mm]
$n$BS  
&  (1\,:\,0\,:\,0)  &  $0\leqq T\leqq\frac{1}{2}$,\,
$\frac{1}{3}\leqq S\leqq 1$,
\tabularnewline[1mm]
 &  & $\frac{1}{2}\leqq T\!+\!S\leqq 1$,\,
 $-1\leqq T\!-\!S\leqq\frac{1}{24}$,\, $\frac{23}{24}\leqq T\!+\!2S\leqq 2$
\\[-3.3mm]
&&
\tabularnewline
\hline
&&
\\[-3.5mm]
Mixture & $(\eta \!:\! 1\!-\!\eta \!:\! 0)$
& $0\leqq (T, S)\leqq 1$,\,
$\frac{1}{2}\leqq T\!+\!S\leqq 1$,\, $\max\{2T\!+\!S,T\!+\!2S\}\geqq \frac{23}{24}$
\\[-3.5mm]
&&
\tabularnewline[0.1cm]
\hline\hline
\end{tabular}
\vspace*{2mm}
\end{table}
\end{center}
%

\vspace*{-8mm}

Recently, the IceCube collaboration\,\cite{IC1} published 37 candidate events
after analyzing its three-year data collection (988 days between 2010\,\textendash\,2013),
with deposited energies within the range of \,30\,\textendash\,2000\,TeV.
Among these events 28 are identified as shower events and 7 as muon-track events.
IceCube also found\,\cite{IC1} that among the 37 recorded events, two events had coincident hits
in the IceTop surface array, so they were almost certainly produced in cosmic ray air showers
and thus should be subtracted.
Although the expected atmospheric background rates
have some uncertainty (e.g., from high-mass mesons with shorter lifetimes),
the energy spectrum, zenith distribution, and muon track to shower ratio of the observed events
strongly disagree with the possibility of having these events from purely atmospheric origin,
at \,$5.7\sigma$\, level.
Hence, these signals should mainly arise from the astrophysical neutrinos with very large $\,L/E$.\,
After subtracting the two atmospheric muon-like events,
the ratio of track events to all signal events is $\,7/35 = 0.2\,$ \cite{IC1},
which indicates that the $\,\nu_\mu^{}\,$ flux ratio $\,T\,$ should be relatively small.
It is worth to note that the value of $\,T\,$ does not necessarily equal
the ratio of track events to total signal events
since the event rate depends on neutrino effective area
which varies for different flavor neutrinos \cite{IC1}.
A recent fit of the flux ratios $(T,\,S)$
by using the three-year IceCube data was given in Ref.\,\cite{ICfit-xi}.\footnote{
It is also worth to note that despite corrections from the neutrino effective areas and
the atmospheric muon backgrounds, technically translating an event topology to the
interacting flavor neutrinos involves complicated analyses.
For example, the High Energy Starting Events (HESE) method\,\cite{IC-rept} requires
the entering particles being energetic enough and is used to select neutrino-like
events by vetoing low energy events in which the earliest light is observed in
the outer part of the detector. The complication of such analysis could induce further
uncertainty for measuring neutrino fluxes and thus affect the value of flavor composition
$\,T\,$.\,  Clearly, a detailed precise determination
of the flux ratios should be eventually done by the experimental collaboration itself.
Besides, since the number of signal events is still small, the value of $\,T$\, is likely to be
subject to changes after more upcoming data are analyzed.}
For the present study, we will compare our general unitarity bounds with the fitted neutrino
flux ratios\,\cite{ICfit-xi},
but we keep in mind that a fully realistic and precise fit to the IceCube
data should be eventually done by the experimental collaboration itself.

In Fig.\,\ref{fig:3}, we further present the recent fit of flux ratios
at \,68\%\,C.L.\,(95\%\,C.L.) \cite{ICfit-xi},
as marked by the red (yellow) shaded area, where the red triangle-dot denotes the best fit.
As we see, if we take Pion Sources with initial flavor ratio $(1:2:0)$,
the fitted flux ratios of IceCube at \,68\%\,C.L.\,(red region) already lie outside of
the unitarity bounds (and the current neutrino global fit\,\cite{fit1}),
but the IceCube constraints at 95\%\,C.L.\,(yellow region) are still consistent with
our unitarity bounds.
Under the unitarity bound $\,S\leqq \frac{1}{2}\,$ from Eq.\,\eqref{eq:120-UB-S},
we find that the 68\%\,C.L.\ fit of IceCube (shaded red area in Fig.\,\ref{fig:3}) restricts
the $\,\nu_\mu^{}\,$ flux ratio $\,T\,$ to a smaller range, $\,T\leqq 0.19$\,.\,
Thus, if future experiments (including IceCube)
could further strengthen this limit and
confirm the source as Pion Source, then new physics would be required to explain a
small flux ratio $\,T\,$ (significantly below $\,23/72\simeq 0.32$\,),\,
such as sterile neutrinos, neutrino decays, pseudo-Dirac neutrinos,
or other exotic effects \cite{exotic}.
The comparison with IceCube in Fig.\,\ref{fig:3} is instructive.
It shows that imposing the unitarity bounds
can put nontrivial universal constraints on the flux ratios $(T,\,S)$.\,

In Fig.\,\ref{fig:3}, the dark spot (consisting of the simulated scattered points) gives the region
allowed by the global fit of current neutrino data \cite{fit1}.
We note that it almost saturates the unitarity bound on the lower left-hand-side of $\,T\,$,
i.e., very close to the unitarity bounds $\,T\geqq 23/72\,$ and $\,T+S\geqq 47/72$\,.\,
This shows that these two unitarity bounds are very important.

For comparison, we also take a canonical reference point
$\,(T,\,S)=(\frac{1}{3},\,\frac{1}{3})$\,,\,
marked as the red point inside the green area of Fig.\,\ref{fig:3},
which corresponds to the flux ratio of \,(1\,:\,1\,:\,1)
at the detector.\footnote{For the initial flavor ratio \,(1\,:\,2\,:\,0)\,
and the neutrino mixing with $\,(\theta_{23}^{},\,\theta_{13}^{})=(\frac{\pi}{4},\,0)$\,,\,
the detected flux ratio would be \,(1\,:\,1\,:\,1)\,.}\,
This point was also discussed before
for the comparison with fitting the IceCube data\,\cite{IC0}\cite{ICfit-xi}.
Fig.\,\ref{fig:3} shows that this red point is excluded by the current neutrino global fit,
and lies nearby the boundary of our unitarity bounds
$\,T+S\geqq 47/72\simeq 0.65\,$ and $\,T\geqq 23/72\simeq 0.32\,$.\,

\vspace*{4mm}
\noindent
\subsection{Muon-Damped Sources with Flavor Ratio (0\,:\,1\,:\,0)}
\vspace*{3mm}

Muon-Damped Sources ($\mu$DS) have an initial flavor ratio (0\,:\,1\,:\,0).\,
Thus, we can infer the flavor ratios at the detector as in \eqref{eq:0423-1},
$\,T=1-X-Z$\, and $\,S=Z$\,.\,
Using the unitarity conditions \eqref{eq:0508-2} and \eqref{eq:0509-2},
we deduce the bounds,
\beqa
\fr{1}{3} \,\leqq\, T \,\leqq\, 1,
&~~~~&
0 \,\leqq\, S \,\leqq\, \fr{1}{2} .
\label{eq:0505-5}
\eeqa

Similar to Sec.\,4.1, for the combinations,
%
\beqa
S+T &\,=\,& 1-X \,,
\label{eq:0505-7}
\nn\\[1.5mm]
2T+S &=& 2-(2X+Z) \,,
\label{eq:0517-3}
\\[1.5mm]
T-S &=& 1-(X+2Z) \,,
\nn\label{eq:0517-4}
\eeqa
%
we derive the following bounds on the flavor ratios,
%
\beqa
\fr{1}{2}\, \leqq\, & \,S\!+\!T\, & \,\leqq\, 1 \,,
\label{eq:0505-8}
\nn\\[1.5mm]
\fr{23}{24}\, \leqq\, & \,2T\!+\!S\, & \,\leqq\, 2 \,,
\label{eq:0505-8-2}
\\[1.5mm]
-\fr{1}{24}\, \leqq\, & \,T\!-\!S\, & \,\leqq\, 1 \,.
\nn\label{eq:0505-8-3}
\eeqa
%

%
\begin{figure}[t]
\centering
\includegraphics[height=9.5cm,width=9.5cm]{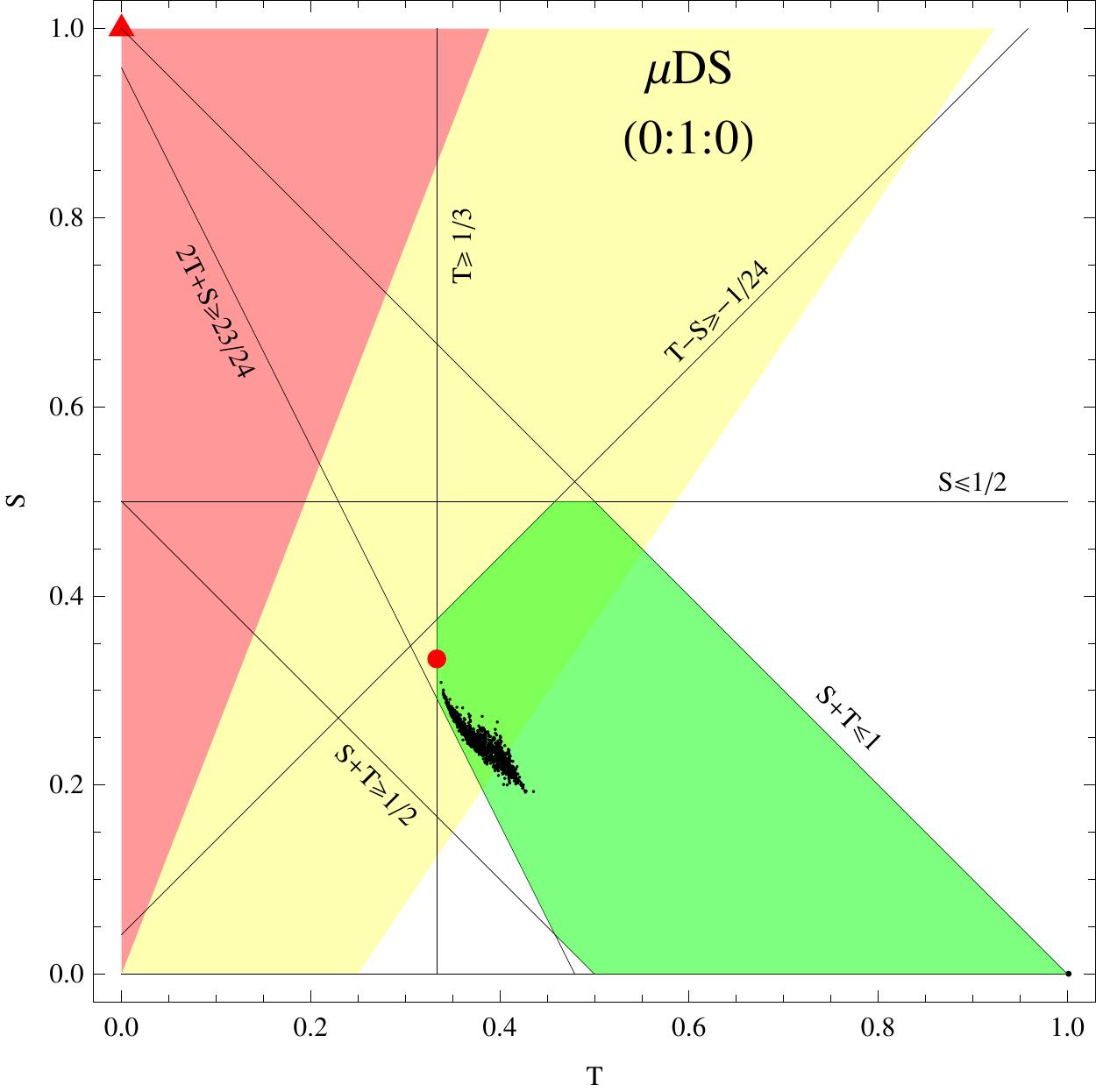}
\vspace*{-2mm}
\caption{Unitarity bounds on the flavor ratios $\,(T,\,S)$,\,
for Muon-Damped Sources ($\mu$DS) with initial neutrino flavor ratio (0\,:\,1\,:\,0).\,
The black straight lines represent the general bounds (\,Table\,\ref{tab:1}\,) derived from
the unitarity of the PMNS matrix without experimental input. These bounds are combined to give
the allowed region (green area). The dark spot inside the shaded area is a collection
of 1000 random points given by the result from a neutrino global fit\,\cite{fit1} of the PMNS matrix.
The red point is defined in the caption of Fig.\,\ref{fig:3}.
The shaded red (yellow) region denotes the fit\,\cite{ICfit-xi} to the three-year IceCube data
at 68\%\,C.L.\,(95\%\,C.L.), with the red-triangle as the best fit.
}
\label{fig:4}
\vspace*{2mm}
\end{figure}

We summarize the above unitarity bounds in Table\,\ref{tab:1} for
$\mu$DS Sources with initial flavor ratio \,(0\,:\,1\,:\,0).\,
We combine these bounds in Fig.\,\ref{fig:4}, and deduce the allowed region of $(T,\,S)$
which is within the shaded green area.
The parameter region allowed by the current neutrino global fit is shown by the black points,
which nearly form a dark spot, same as in Fig.\,\ref{fig:3}.
We note that in this case the dark spot region
almost saturate the unitarity bounds on $\,T\,$ from its lower side.
This shows that the general bounds $\,T\geqq \frac{1}{3}\,$ and
$\,2T+S\geqq \frac{23}{24}\,$ play an important role here.
Furthermore, Fig.\,\ref{fig:4} shows that,
if we take $\mu$DS as the neutrino sources with initial flavor composition \,(0\,:\,1\,:\,0),\,
the current fit\,\cite{ICfit-xi} to the three-year IceCube data lies outside of
the unitarity bound (green region) at 68\%\,C.L.\,(red region),
but is still consistent with the unitarity constraint (green region) at $\,95\%\,$C.L.\,(yellow region).
Here we would note again that a fully realistic fit to the flavor compositions
should be done by the experimental collaboration.

\pagebreak
\vspace*{0.3mm}
\noindent
\subsection{Neutron Beam Sources with Flavor Ratio (1\,:\,0\,:\,0)}
\vspace*{3mm}

\begin{figure}[t]
\centering
\includegraphics[height=9.5cm,width=9.5cm]{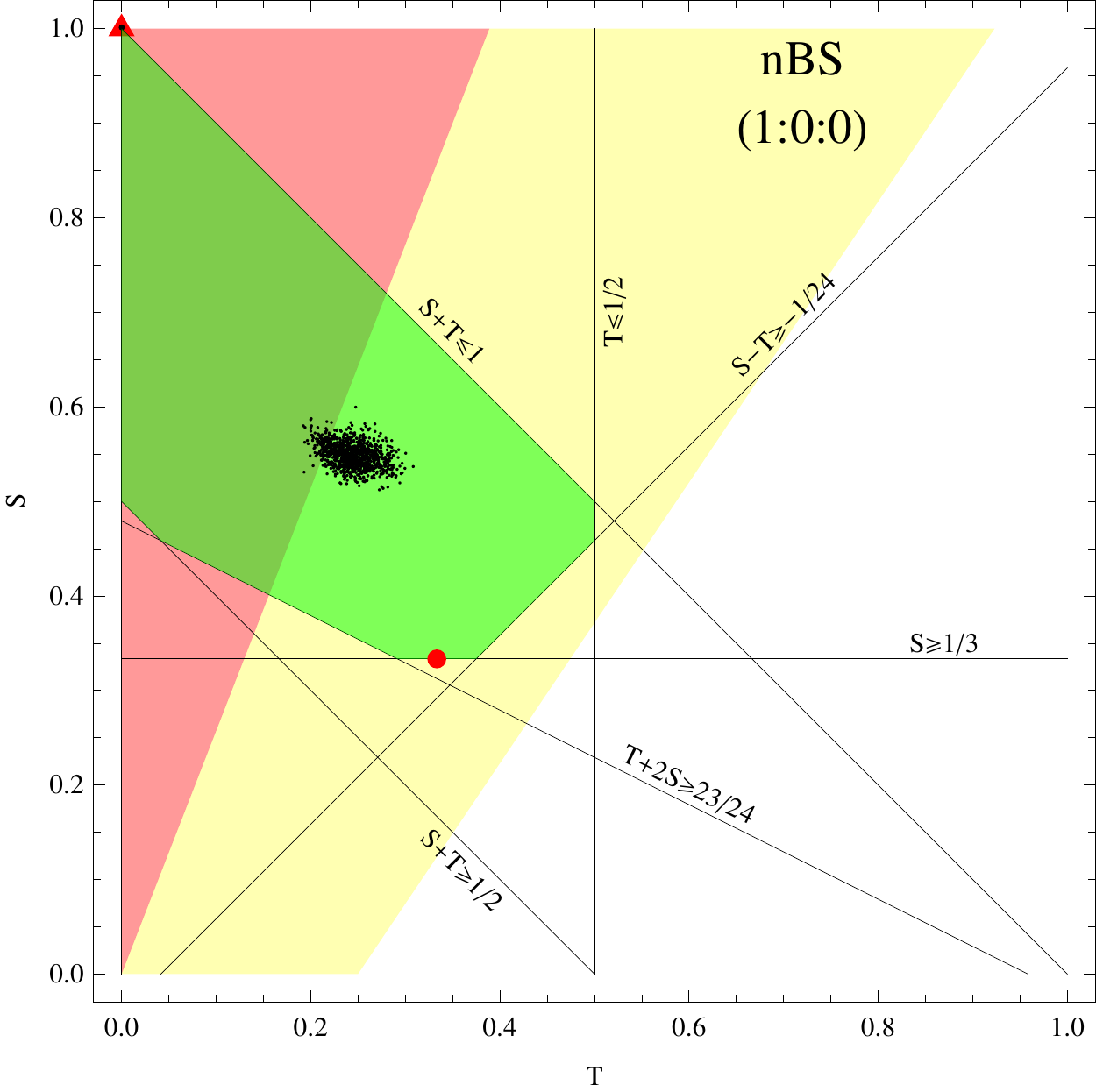}
\caption{Unitarity bounds on the flavor ratios $(T,\,S)$,\,
for Neutron Beam Sources (nBS) with initial flavor ratio \,(1\,:\,0\,:\,0)\,.\,
The black straight lines represent the general bounds (\,Table\,\ref{tab:1}\,) derived from the unitarity
of PMNS matrix without experimental input. These bounds are combined to give the allowed region
as shown by the shaded green area (including the overlapping part).
The dark spot inside the shaded area is a collection
of 1000 random points given by the current neutrino global fit\,\cite{fit1} of the PMNS matrix.
The red point is defined in the caption of Fig.\,\ref{fig:3}.
The shaded red (yellow) region denotes the fit\,\cite{ICfit-xi} to the three-year IceCube data
at 68\%\,C.L.\,(95\%\,C.L.), with the red-triangle as the best fit.
This fit is fully consistent with the unitarity bounds and
the current global fit\,\cite{fit1} of the PMNS matrix.
}
\label{fig:5}
\end{figure}

For astrophysical neutrinos from Neutron Beam Sources (nBS) with initial flavor ratio
\,(1\,:\,0\,:\,0)\,,\, our unitarity analysis is similar to that of Sec.\,4.2
for $\mu$DS with the initial flavor ratio \,(0\,:\,1\,:\,0)\,.\,

In the case of nBS sources, we have
$\,T=Z\,$ and $\,S=1-Z-Y\,$.\,  Thus, we derive the combinations
$S+T$, $T+2S$, and $S-T$ in terms of $\,(X,\,Y,\,Z)$\,,
%
\beqa
S+T &\,=\,& 1-Y \,,
\label{eq:0505-7-1}
\nn\\[1.5mm]
T+2S &=& 2-(2Y+Z) \,,
\label{eq:0517-3-1}
\\[1.5mm]
S-T &=& 1-(Y+2Z) \,.
\nn\label{eq:0517-4-1}
\eeqa
%

In parallel to Sec.\,4.2, we derive unitarity bounds on the flavor ratios,
\beqa
0 \,\leqq \,T\, \leqq\, \fr{1}{2},~~~~~
\fr{1}{3} \,\leqq \,S\, \leqq\, 1 \,, ~~~~
\label{eq:0505-5-1}
\eeqa
and their combinations above,
\beqa
\fr{1}{2} \,\leqq &~S\!+\!T~& \leqq 1\,,
\label{eq:0505-8-1}
\nn\\[1.5mm]
\fr{23}{24} \,\leqq &~T\!+\!2S~& \leqq\, 2 \,,
\label{eq:0505-8-2-1}
\\[1.5mm]
-\fr{1}{24} \,\leqq &\,S\!-\!T\,& \leqq\, 1 \,.
\nn\label{eq:0505-8-3-1}
\eeqa
%

We summarize these bounds into Table\,\ref{tab:1},
and present their combined bounds (shaded green area) in Fig.\,\ref{fig:5}.
From this plot, it is interesting to see that the current fit\,\cite{ICfit-xi} to the IceCube data
is compatible with the unitarity bounds (green area) for the nBS sources
already at 68\%\,C.L.\,(red area), because of the large overlapping region.
Furthermore, at 95\%\,C.L.\,(yellow area) the fit\,\cite{ICfit-xi} is fully consistent with
the unitarity bounds, as well as the current neutrino global fit\,\cite{fit1}
(as represented by the black points in the dark spot region).
Comparing Fig.\,\ref{fig:5} with Fig.\,\ref{fig:3}-\ref{fig:4}, we see that the fit\,\cite{ICfit-xi}
tends to favor the origin of astrophysical neutrinos to contain a sizable fraction of the nBS sources
with initial flavor composition \,(1\,:\,0\,:\,0).\,
In the next subsection, we will further analyze a mixed source
of the three types, with a generic flux ratio $(\eta:1\!-\!\eta:0)$.

\vspace*{3mm}
\noindent
\subsection{Mixed Sources with Flavor Ratio $(\eta\!:\!1\!-\!\eta\!:\!0)$}
\label{sec:4.4}
\vspace*{1.5mm}

\begin{figure}[t]
\centering
\includegraphics[width=9.5cm,height=9.5cm]{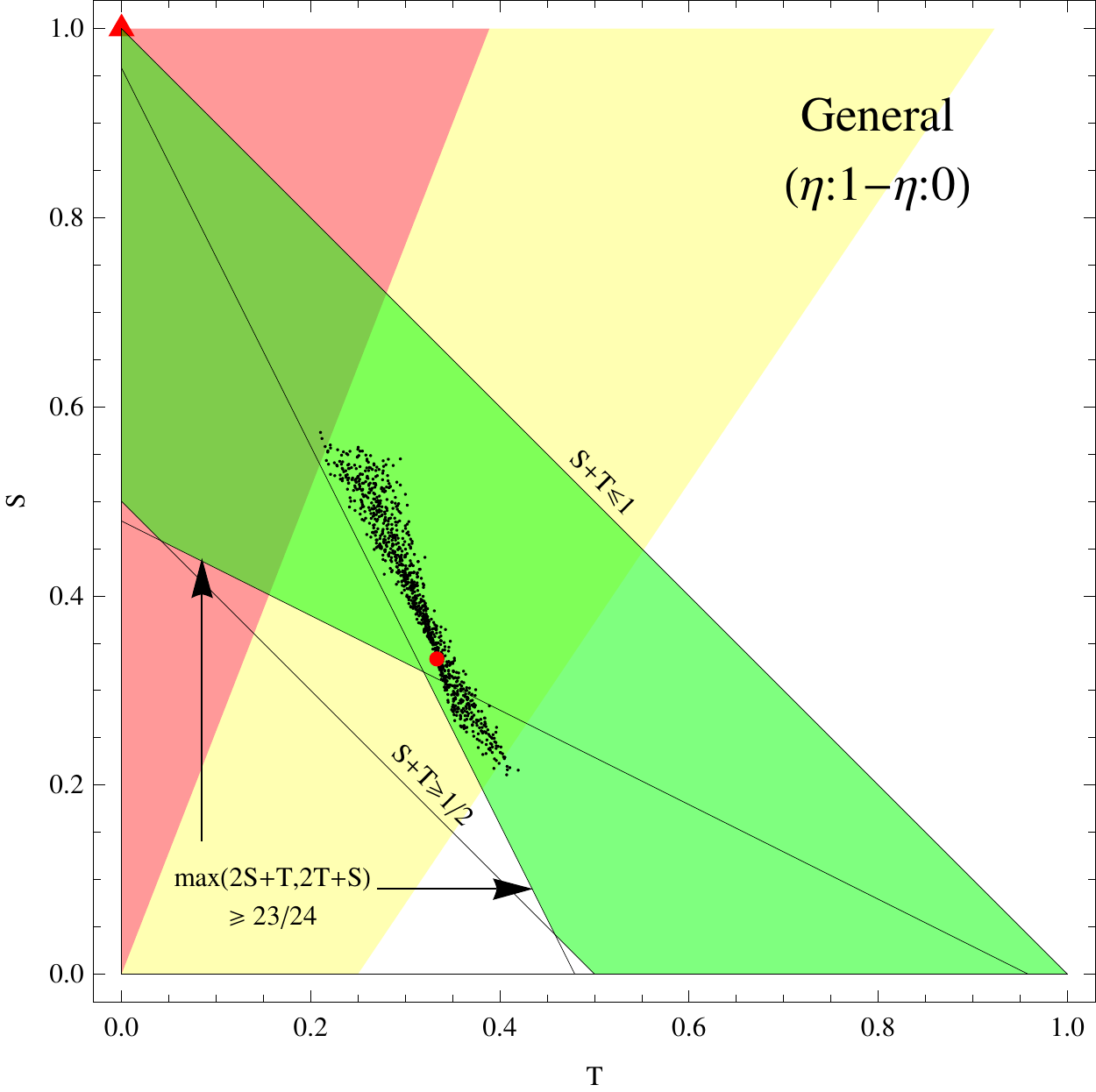}
\vspace*{-2mm}
\caption{Unitarity bounds on the flavor ratios $(T,\,S)$,\,
for a generic mixture of three types of sources ($\pi$S, $\mu$DS, $n$BS)
with initial flavor ratio $\,(\eta:1\!-\!\eta:0)\,$ and $\,\eta\in [0,\,1]$.\,
The black straight lines represent the general bounds (\,Table\,\ref{tab:1}\,)
derived from the unitarity of PMNS matrix without experimental input.
These bounds are combined to give the allowed green region (including the overlapping part).
The dark spot inside the shaded area collects 2000 random points
given by the current neutrino global fit of PMNS matrix and with a scan of $\,\eta\in [0,\,1]\,$.\,
The red point is defined in the caption of Fig.\,\ref{fig:3}.
The shaded red (yellow) region denotes the fit\,\cite{ICfit-xi} to the three-year IceCube data
at 68\%\,C.L.\,(95\%\,C.L.), with the red-triangle as the best fit.
This fit is consistent with the unitarity bounds.
}
\label{fig:6}
\vspace*{4mm}
\end{figure}

In general, we can consider a mixed neutrino source of the three types above,
where the $\,\nu_{\tau}^{}\,$ neutrinos are absent. So, this general source has
the initial flavor ratio  $\,(\eta:1\!-\!\eta:0)\,$ with $\,\eta\in [0,\,1]$.\,
In this notation, Pion Sources correspond to $\,\eta = \frac{1}{3}\,$,\, Neutron Beam Sources
to $\,\eta = 1\,$,\, and Muon-Damped Sources to $\,\eta = 0$\,.\,
Thus, for the general case we have the flavor ratios,
\beqs
\label{eq:mix-TS}
\begin{eqnarray}
T & \,=\, & \eta Z+(1\!-\!\eta )(1\!-\!X\!-\!Z) \,,
\label{eq:0423-1-2}
\\[1.5mm]
S & \,=\, & \eta (1\!-\!Z\!-\!Y)+(1\!-\!\eta )Z \,.
\label{eq:0505-10}
\end{eqnarray}
\eeqs
As we noted below Eq.\,\eqref{eq:0602-2}, under the exchange
$\,\nu_e^{}\leftrightarrow\nu_\mu^{}\,$,\, we have,
$\,(X,\,Z)\leftrightarrow (Y,\,Z)\,$ and $\,\eta\leftrightarrow (1\!-\!\eta)\,$.\,
Then, from \eqref{eq:mix-TS}, we see that this exchange
leads to $\,S\leftrightarrow T\,$.\,
This property will also ensure the unitarity bound (the shaded region with light blue color)
in Fig.\,\ref{fig:6} to be symmetric with respect to the line $\,S=T\,$.\,

For a given $\,\eta\,$ and $\,T\,$,\, we may view (\ref{eq:0423-1-2}) as
a straight line in the $X\!-\!Z$ plane of Fig.\,\ref{fig:2},
\beqa
Z \,=\,
-\frac{1\!-\!\eta}{\,1\!-\!2\eta\,}X
-\frac{\,T\!-\!(1\!-\!\eta)\,}{\,1\!-\!2\eta\,}  \,,
\label{eq:line-ZX-T}
\eeqa
where the slope is fully determined by $\,\eta\,$,\,
and $\,T\,$ only affects the intercept at $\,X=0\,$.\,
In the $XYZ$ coordinate frame of Fig.\,\ref{fig:2}, Eq.\,\eqref{eq:line-ZX-T} describes
a plane which is perpendicular to the $X\!-Z$ plane and intersects with it at the line
given by \eqref{eq:line-ZX-T}.
It is clear that for $\,\eta\in [0,\,1/2)\,$ and as $\,T\,$ decreases,
the plane \eqref{eq:line-ZX-T} increases its intercept
at $Z$ axis in Fig.\,\ref{fig:2}. If $\,T$ decreases to a value
such that this plane no longer intersects the space surrounded
by the colored surfaces in Fig.\,\ref{fig:2}, i.e., every point
in this plane violates the unitarity bound, then this value of $\,T\,$
is disallowed by the unitarity.  Thus, we can derive a lower bound on $\,T\,$
by moving the plane \eqref{eq:line-ZX-T} to the ``critical position" where it is just
going to fully leave the colored surfaces (unitarity bounds) of Fig.\,\ref{fig:2}.

For $\,\eta\in [0,\,1/3]$,\, the slope of the line \eqref{eq:line-ZX-T}
is restricted within $\,[-2,\,-1]\,$.\,
So, from Fig.\,\ref{fig:2}, its critical position should be the point B in the $X-Z$ plane,
and has the coordinates $(X,\,Z)=(3/8,\,7/24)$.\, At the point B,
Eq.\,(\ref{eq:0423-1-2}) gives, $\,T[B]=(8-\eta )/24\,$.\,
This is gives the lower bound $\,T\geqq (8-\eta )/24\,$ for $\,\eta\in [0,\,1/3]$.\,
Similarly, for $\,\eta\in [1/3,\,1/2]\,$,\, the slope of \eqref{eq:line-ZX-T}
is within $\,(-\infty ,\,-2]$.\,  We find that the critical position of \eqref{eq:line-ZX-T}
should be the point $A$ and has the coordinates $(X,\,Z)=(1/2,\,1/24)$.\,
Thus, we compute the value of $\,T\,$ at this point, $\,T[A]=(11-10\eta)/24\,$.\,
Hence, we deduce the unitarity bound $\,T\geqq (11-10\eta )/24\,$ for $\,\eta\in [1/3,\,1/2]$.\,
Finally, for $\,\eta\in [1/2,\,1]$,\, the slope of \eqref{eq:line-ZX-T}
is within $\,[0,\,+\infty)$.\, In this case, the intercept of \eqref{eq:line-ZX-T} is the sum
of two parts, $\,(1-\eta)/(1-2\eta)\in (-\infty,\, 0]\,$ and  $\,T/(2\eta -1)>0\,$.\,
Thus, decreasing $\,T\,$ will reduce the intercept and move the line \eqref{eq:line-ZX-T} downwards,
which will reach a critical position at the point $\,(X,\,Z)=(1/2,\,0)\,$.\,
At this point we find the corresponding lower bound, $\,T\geqq (1-\eta)/2\,$.\,
Combining these bounds, we arrive at
\beq
\ba{lcll}
T & \geqq & \fr{1}{24}\(8 - \eta\) \,,
&~~ \textrm{for}~ \eta \in \left[0,\,\frac{1}{3}\right],
\\[2mm]
T & \geqq & \fr{1}{24}\(11 - 10\eta\) \,,
&~~ \textrm{for}~ \eta \in \left[\frac{1}{3},\,\frac{1}{2}\right],
\\[2mm]
T & \geqq & \dis\fr{1}{2}\(1 - \eta\) \,,
&~~ \textrm{for}~ \eta \in \left[\frac{1}{2},\,1\right].
\ea
\label{eq:0518-1}
\eeq

From Eq.\,\eqref{eq:0518-1}, we note that for $\,\eta \leqq \frac{1}{2}$,\,
$\,T \geqq \frac{1}{4}\,$ always holds; and
only when $\,\eta > \frac{1}{2}$\,,\, we have $\,T < \frac{1}{4}\,$.\,
Hence, even if we do not know which types of cosmic neutrino sources are invoked,
from the IceCube measurement we can extract important information about the initial
flavor composition. For instance, if the measured $\,T\,$ value by IceCube confirms
$\,T < \frac{1}{4}\,$  and there is no $\,\nu_{\tau}^{}\,$ source,
then $\,\eta >\frac{1}{2}\,$ has to hold.
Hence, we can infer that more $\nu_{e}^{}$ neutrinos
than $\nu_{\mu}^{}$ neutrinos exist in the initial flavor composition of cosmic neutrinos.

Next, we further analyze the unitarity constraints on $\,T\!+\!S\,$.\,
From \eqref{eq:mix-TS}, we have
\beqa
T+S \,=\, 1-\left[\,\eta Y+(1-\eta)X\,\right] .
\label{eq:0517-7}
\eeqa
From Eqs.\,\eqref{eq:0508-1} and \eqref{eq:0508-2},
we have $\,0\leqq (X,\,Y)\leqq\frac{1}{2}\,$.\,
Thus, for $\,\eta\in [0,\,1]\,$, we can deduce
\beqa
\frac{1}{2} \,\leqq\, T\!+\!S \,\leqq\, 1 \,,
\label{eq:0505-8-1-1}
\eeqa
where the lower bound is reached for $\,X=Y=\frac{1}{2}$\,
and the upper bound is saturated if $\,X=Y=0\,$.\,

With the formulas from Eq.\,\eqref{eq:mix-TS},
we deduce the combinations $\,2T\!+\!S$\, and $\,T\!+\!2S$\, as follows,
\beqs
\label{eq:2TS-2ST}
\beqa
\hspace*{-6mm}
2T\!+\!S &=& (2\!-\!\eta) \!-\! \left[2(1\!-\!\eta)X \!+\!\eta Y \!+\!(1\!-\!2\eta)Z\right]
\label{eq:eta-2TS}
\\[1.5mm]
\hspace*{-6mm}
&=& (2\!-\!\eta) \!-\! \left[\eta (2X\!+\!Y)\!+\!(1\!-\!2\eta)(2X\!+\!Z)\right],~~~
\nn\\[2mm]
\hspace*{-6mm}
T\!+\!2S &=& (1\!+\!\eta) \!-\!
\left[(1\!-\!\eta)X \!+\! 2\eta Y\!-\! (1\!-\!2\eta)Z\right]
\label{eq:eta-2ST}
\\[1.5mm]
\hspace*{-6mm}
&=& (1\!+\!\eta) \!-\! \left[(1\!-\!\eta)(2Y\!\!+\!X)\!+\!(2\eta\!-\!1)(2Y\!\!+\!Z)\right].
\nn
\eeqa
\eeqs
As a consistency check, we note that the above formula for $\,T\!+\!2S\,$ can be
inferred from $\,2T\!+\!S$\, by the exchanges of $\,(X,\,Z)\leftrightarrow (Y,\,Z)$\,
and $\,\eta\leftrightarrow (1\!-\!\eta)$,\,
because these exchanges lead to $\,S\leftrightarrow T\,$ and thus
$\,(2T\!+\!S)\leftrightarrow (T\!+\!2S)\,$.\,

Using the conditions \eqref{eq:bound-Y+2Z},
we derive the following lower bounds from \eqref{eq:eta-2TS} with
$\,\eta\leqq\frac{1}{2}\,$ and from \eqref{eq:eta-2ST} with $\,\eta\geqq\frac{1}{2}\,$,
respectively,
\beq
\ba{lcll}
2T\!+\!S   
& \geqq & \dis \frac{\,23+\eta\,}{24}\,, &~~
\textrm{for~} \eta \in \left[0,\,\frac{1}{2}\right], 
\\[3mm]
T\!+\!2S   
& \geqq & \dis 1-\frac{\,\eta\,}{24}\,, &~~
\textrm{for~} \eta \in \left[\frac{1}{2},\,1\right].
\ea
\label{eq:2TS}
\eeq
For any $\,\eta\in [0,\,1]\,$,\, we see at least one of the combinations
$\,2T\!+\!S$\, and $\,T\!+\!2S$\,
is larger than $\,\frac{23}{24}\,$.\, Hence, we have the combined lower bound,
\begin{equation}
\max\{2T\!+\!S,\,2S\!+\!T\}\,\geqq\, \frac{23}{24}\,.
\hspace*{10mm}
\label{eq:0505-9}
\end{equation}

From the above analysis, we summarize the unitarity constraints
\eqref{eq:0505-8-1-1} and \eqref{eq:0505-9}
for the generic flavor ratio $(\eta:1\!-\!\eta:0)$ in Table\,\ref{tab:1},
which hold for any $\,\eta\in [0,\,1]\,$.\,
For demonstration, we further present these general bounds in
Fig.\,\ref{fig:6}, where we derive the combined unitary bound in
the $\,T-S\,$ plane, as depicted by the green area (including the overlapping part).
As we expected earlier, this unitarity bound (green region) is symmetric
respect to the line $\,S=T\,$.\,
We note that this general bound holds for any $\,\eta\,$ value and
is weaker than the bounds of Figs.\,\ref{fig:3}$-$\ref{fig:5}
(which correspond to specific $\,\eta\,$ values).
Actually, each bounded area in Figs.\,\ref{fig:3}$-$\ref{fig:5} is contained
as a certain part of the allowed region in Fig.\,\ref{fig:6}.

\vspace*{1mm}

From the general bounds in Fig.\,\ref{fig:6}, we see that
even though the type of the cosmic neutrino sources is unknown a priori,
we can still deduce nontrivial unitarity constraints on the flux ratios $\,T\,$ and $\,S\,$.\,
This means that for any source among ($\pi$S,\,$\mu$DS,\,$n$BS) or their
general mixture in its initial flavor composition,
the flux ratios $\,(T,\,S)$\, must lie in the shaded green region
of Fig.\,\ref{fig:6}. Otherwise, the unitarity bounds are violated, which would require proper
underlying new physics. For a comparison, we present the recent fit\,\cite{ICfit-xi}
to the three-year IceCube data in the same plot, shown as the shaded red (yellow) area
at 68\%\,C.L.\,(95\%\,C.L.).
The plots in Fig.\,\ref{fig:5}$-$\ref{fig:6}
also suggest that more precise measurements of the $\,\nu_e^{}\,$ flux ratio $\,S\,$
will be important for pinning down the initial flavor composition in the source.

\vspace*{2mm}
\section{Conclusions}
\label{sec:5}
\vspace*{2mm}

Observations of ultra-high-energy astrophysical neutrinos at IceCube\,\cite{IC0,IC1}
have marked the exciting start of neutrino astronomy.  This may eventually
help astronomers to map individual sources of astrophysical neutrinos in the sky,
and thus paint a picture of the universe by means of neutrino telescopes.

\vspace*{1mm}

In this work, we made use of the unitarity of leptonic PMNS mixing matrix,
and systematically derived unitarity constraints on the flavor composition of astrophysical neutrinos,
in comparison with the recent fit\,\cite{ICfit-xi} to the three-year IceCube data\,\cite{IC1}
and the current neutrino global fit\,\cite{fit1,fit2}.
In Section\,\ref{sec:2}, using the leptonic unitarity triangles (LUTs) \cite{He:2013rba}, 
we formulated the flavor transition probabilities of astrophysical neutrinos 
in terms of the geometrical parameters of the LUTs, as given in
Eqs.\,\eqref{eq:P-abc}$-$\eqref{eq:P-dis}. Then, we expressed the $\nu_\mu^{}$ and $\nu_e^{}$
flux ratios $(T,\,S)$ by the LUT parameters in Eqs.\,\eqref{eq:0423}$-$\eqref{eq:TS-XYZ-eta}
for different neutrino sources and their mixture.
In Section\,\ref{sec:3},
we quantitatively derived nontrivial unitarity bounds on the transition probabilities of
cosmic neutrinos by using the geometrical conditions (such as the triangular inequalities).
These are presented in Eqs.\,\eqref{eq:0508-2}$-$\eqref{eq:0508-4} and
Eqs.\,\eqref{eq:0509-2}$-$\eqref{eq:bound-Y+2Z}, as well as Figs.\,\ref{fig:1}$-$\ref{fig:2}.
These and other new bounds we derived generally hold for three flavor neutrinos,
independent of any experimental input or the pattern of leptonic mixing.

\vspace*{1mm}

In Section\,\ref{sec:4}, we applied these generic unitarity bounds to impose constraints on the
flux ratios \,$(T,\,S)$\, for three types of the neutrino sources ($\pi$S, $\mu$DS, $n$BS)
and their general mixture. These unitarity constraints are summarized in Table\,\ref{tab:1}.
In Figs.\,\ref{fig:3}$-$\ref{fig:6}, we compared these constraints with the IceCube data\,\cite{IC1},
as well as the current neutrino global fit\,\cite{fit1}.
With the cosmic neutrino flux ratios extracted from the recent fit\,\cite{ICfit-xi} to the
three-year IceCube data\,\cite{IC1}, we found that the $\pi$S or $\mu$DS sources
would be disfavored by unitarity bounds at 68\%\,C.L., but still consistent with
the unitarity at 95\%\,C.L., as shown in Figs.\,\ref{fig:3}$-$\ref{fig:4}.
If the $\pi$S or $\mu$DS sources are the real origin of the observed astrophysical neutrinos,
a further confirmation of the IceCube data away from the unitarity bounds
would suggest either a misidentification of certain tracks as showers,
or a misunderstanding of the potential background events,
or the existence of underlying new physics beyond the standard model.
In addition, for neutrino sources such as the $n$BS or the mixed sources,
we revealed that the recent IceCube fit\,\cite{ICfit-xi} is consistent with our unitarity bounds
(as well as the current neutrino global fit) at both 68\%\,C.L.\ and 95\%\,C.L.,
as shown in Figs.\,\ref{fig:5}$-$\ref{fig:6}.
Even without specifying the type of sources, the suggested flavor ratio \,(1\,:\,1\,:\,1)\,
at the detector is within and very close to our unitarity bound,
and is compatible with the IceCube data at 95\%\,C.L.\ (Fig.\,\ref{fig:6}).\footnote{As we commented
in footnote-2, the complication of uncertainties of the experimental HESE analysis may cause possible
mis-reconstruction of events at the topology level.
Hence, a fully realistic and reliable fit to the IceCube data should be done by the
experimental collaboration itself, which will put any conclusion on a firmer ground.}~

\vspace*{1mm}

Finally, in Section\,\ref{sec:4.4},
we proved that for any sources without $\nu_\tau^{}$ neutrinos (such as $\pi$S,
$\mu$DS, $n$BS, or their mixture), a detected $\nu_\mu^{}$ flux ratio $\,T < {1}/{4}\,$
will require the initial flavor composition with more $\,\nu_e^{}\,$ neutrinos than
$\,\nu_\mu^{}\,$ neutrinos.

\vspace*{6mm}
\noindent
{\bf\large Acknowledgements}\\[1.5mm]
We thank John R.\ Ellis, Eligio Lisi, Sergio Palomares-Ruiz and Xiangyang Yu for useful discussions.
XJX and HJH were supported by Chinese NSF (Grant No.\ 11275101 and 11135003)
and National Basic Research Program (No.\ 2010CB833000).
WR was supported by the Max Planck Society in the project MANITOP.

\vspace{3mm}
%

\end{document}